# Student Use of a Single Lecture Video in a Flipped Introductory Mechanics Course

John M. Aiken[*], Shih-Yin Lin[†], Scott S. Douglas[*], Edwin F. Greco[*], Brian D. Thoms[¶], Marcos D. Caballero[$] and Michael F. Schatz[*]

*School of Physics, Georgia Institute of Technology, 830 State Street, Atlanta, GA 30332
† Department of Physics, National Changhua University of Education, Changhua, Taiwan 500
¶ Department of Physics and Astronomy, Georgia State University, Atlanta, GA 30303
$Department of Physics and Astronomy, Michigan State University, East Lansing, MI 48824

**Abstract:** In the Fall of 2013, Georgia Tech offered a "flipped" calculus-based introductory mechanics class as an alternative to the traditional large-enrollment lecture class. This class flipped instruction by introducing new material outside of the classroom through pre-recorded, lecture videos. Video lectures constituted students' initial introduction to course material. We analyze how students engaged with online lecture videos via "clickstream" data, consisting of time-stamped interactions (plays, pauses, seeks, etc.) with the online video player. Analysis of these events has shown that students may be focusing on elements of the video that facilitate a "correct" solution.

## INTRODUCTION

Beginning in the Fall of 2013 Georgia Tech offered a "flipped" version of their Introductory Mechanics course taught using the Matter and Interactions (M&I) curriculum[1,2]. This "flipped" course introduced new content to Georgia Tech students via lecture videos hosted on the Coursera Massive Open Online Course (MOOC) platform. Using the interaction logs for a single video (i.e., "clickstream" data), we describe how Georgia Tech students interacted with the online video player. Such descriptions promise to give us a window into how students engage material outside of class, the time they spend on these tasks, and what content should be highlighted during class.

## Flipping an Introductory Mechanics Course

A single section of an M&I mechanics course (N=161) at Georgia Tech was first flipped in Fall 2013. Our intention in flipping this class was to spend more class time on group problem solving, rather than on lecturing. The live lecture was replaced by lecture videos originally created for a MOOC version of the course that has been offered since the Summer of 2013[3]. These lecture videos introduced students to mechanics concepts (forces, momentum, energy, etc.) and laboratory practices (programming in python, analyzing video using motion capture software, etc.).

Laboratory activities were also moved outside the classroom. Individually, students observe motion in the world around them using tools in their own environment: smartphone cameras and personal computers. Laboratory activities are built around students capturing and analyzing motion in their environment[4,5]. Additionally, students reported on their findings using video lab reports that are peer-evaluated. In addition to engaging students in scientific practice, these activities were designed to teach students how to think critically about scientific information that they encounter in the real world and how to communicate and evaluate scientific arguments.

Labs were central to this course. A number of lecture videos were intended to teach practices and concepts necessary for successfully engaging with the lab activities. In this paper, we analyze student interaction with a single video that introduces students to the use of Python to develop predictive models of motion with Newton's 2$^{nd}$ Law. By characterizing the content of the parts of the video to which students attend, we can gain insight into how students use video resources when performing these lab activities.

## CLICKSTREAM

"Clickstream" data has been used almost since the advent of the Internet to analyze how users interact with web content[6]. The "stream" is a time-stamped log of each interaction a user makes with a computer application. In this case, we are looking at how users interact with the video player hosted by Coursera.

## Video in a Flipped Class

Videos hosted by Coursera can be viewed like YouTube videos. Students can pause, play, fast forward, and rewind videos, and can change the rate of play. Videos on Coursera can also include instructor-created "interaction points". An interaction point automatically pauses the video, and prompts the student to click a web link or answer a question. The video analyzed in this paper, "Creating a Computer Model of Constant Velocity Motion" (http://youtu.be/DcjIgJY5ets), has five such interaction points. The first interaction point referred students to a video emphasizing the importance of learning how to program (http://youtu.be/nKIu9yen5nc). The second interaction point linked students to the files needed to complete the activity. The next three interaction points posed conceptual questions to the students about the Python code.

The video analyzed in this paper was the first introduction students had to using Python to model physical phenomena. This video was designed to teach students how to create a computational model based on Newton's 2$^{nd}$ law to describe/predict the observed constant velocity motion of a ball. The video discussed the importance of computation to science, walked through a python script that students used to model the motion of a ball moving at constant velocity, and had a series of interaction points posing conceptual questions about the physical and mathematical meaning of the python code. Variable assignment, iteration, commenting and other programming concepts were also introduced in this video.

The video further emphasized starting a computational model from Newton's 2$^{nd}$ law first[7]. Thus, every student should have implemented a Newton's 2$^{nd}$ law update ($v_f = v_i + \frac{F_{net}}{m} dt$) in their Python code, even though the net force was zero. In order to help students follow along with the video, every student was given the same observational data and they were encouraged to practice creating the model themselves as they watched the video.

We chose to analyze this video because it had a large number of views (N=622) and introduced many important programming and physics concepts used repeatedly throughout the course. Since the video was designed as a "walkthrough" for the lab activity, it gave students a chance to engage in lab practices (in this case, building a computational model of real-world data) before they employed them for the actual lab activity.

## Analyzing a Single Student's Interaction

To provide a sense of how our analysis of all views was conducted, we present a detailed analysis of a single viewing by a single student. Figure 1 shows the third time a single student viewed this video; this was the first time the student watched the video to its conclusion. This type of viewing behavior was common for this video. Students' first views of this video often terminated at the second interaction point (out of five total) because at this point students would download the starter python script.

There are three types of events recorded of student interaction that are analyzed in this paper: plays, pauses, and seeks. Plays (represented by green triangles in Figure 1) and pauses (represented by red circles in Figure 1) can be auto-generated by the video player or manually generated by the student. There will always be a play at the beginning of each video and a pause at the end. Interaction points (represented by horizontal dashed lines in Figure 1) auto-generate pauses if the video player reaches the interaction point. In Figure 1, the student sought past the first interaction point and thus there is no auto-generated pause.

The blue diagonal line in Figure 1 represents where "video time" and "real time" are identical (the video is 1052 seconds long, or ~17.5 minutes). Thus, events appearing above the line happened faster than playback (assuming a playback rate of 1.0x) and events appearing below the line happened slower than playback.

For each student-video interaction, plots like Figure 1 provide a "watching trajectory", that is, a snapshot of how each student interacted with the video. Beginning from the origin of Figure 1, this student watched the video for approximately 32 seconds in real-time and then began to seek through the video. Because the video did not pause at the first interaction point (represented by the first dashed line), this provides evidence that the student bypassed this interaction point. The video auto-paused at the second interaction point, and playback was resumed ~2.7 seconds later. The student then watched the video for ~370 seconds and then "seeked to pause" at time 528 s in the video (here, the initial conditions for the computational model were discussed explicitly). The student then watched the video for another ~82 seconds until pausing for ~64 seconds at time 610 s in the video (here, the iterative time step was discussed). Note that this pause caused the total "interaction time" to exceed the total play time of the video (video length = 1052 s, video interaction time = 1200 s). Subsequent pause-play pairs corresponded to remaining three interaction points in the video.

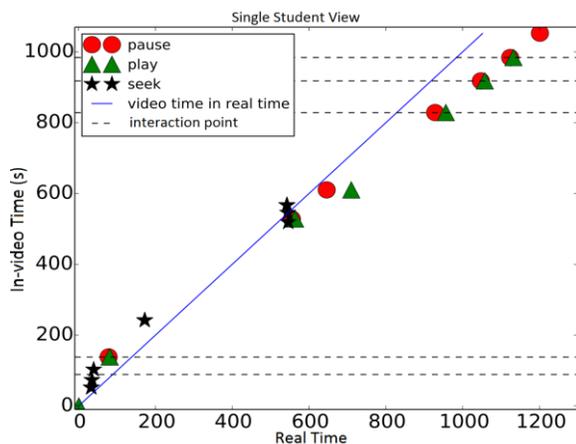

**Figure 1** This is the third view (and first complete view) of "Creating a Computer Model of Constant Velocity Motion" by a single student. The blue diagonal line represents where video time and real time are identical. The dashed lines indicate interaction points

From this watching trajectory, we conclude that this student focused on the finding specific values that were necessary to complete the assignment (noted by the seeking and pausing in the middle of the video). This student spent a total of 46.8 s on the conceptual final three interaction point questions. The third interaction point (27.9 s time on task) was particularly important to the completion of the lab; it asked the students to choose the correct code statement from a list representing Newton's $2^{nd}$ law correctly.

## Analyzing All Students' Interactions

Students viewed "Creating a Computer Model of Constant Velocity Motion" 622 times during the Fall 2013 semester (4.2±0.2 average views per student). 148 students (91.9% of the class) viewed the video at least once. Students' initial viewing behavior differed from their subsequent viewing behavior. We believe this is due to the nature of the video: the second interaction point had a link to download the starter python script. Students may have downloaded the code and began working on their lab assignment, returning to the video later. Clicking the download link does not change the URL of the active page, it opens a new page, and thus students were choosing to ignore the video after clicking the link. Students had the ability to play, pause, and seek through the video at their leisure. Figure 2 shows the pause and seek events for all views of the "Creating a Computer Model of Constant Velocity Motion". Since interaction points (labeled in Figure 2 by blue dotted vertical lines) auto-generate pause events, and since we are interested in students' deliberate interactions, pauses with "in-video" timestamps coinciding with interaction point timestamps have been removed. Video buffering issues, browser idiosyncrasies, and other issues can cause pauses to auto-generate as well. Thus a "noise" measurement was developed using the median absolute deviation (MAD)[8]. The MAD was used because the distribution of pauses is not normal ($\chi^2$=116.7, p<<0.01) and the MAD is insensitive to outliers. In Figure 2, the noise has been subtracted out using two median absolute deviations (2σ) above the median. The color codes at the top of Figure 2 correspond to the content found in the video. These color codes make up the "video timeline" discussed in the next section.

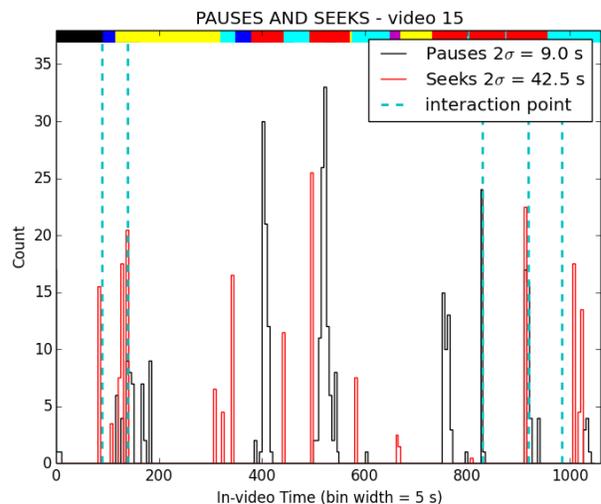

**Figure 2** The color codes at the top of the graph correspond to the timeline codes defined in Table 1. The pause and seek distributions have been shifted by subtracting two MADs from the median. Student pauses (black) appear to focus on "Parameter/Code Values" (red color code).

*Video Timeline Categories*

The video timeline was created to provide a graphical representation of what is going on in the video itself. The timeline is not a specific description of each second of the video. Instead, it attempts to use a standard set of categories to describe generally what is happening in the video. Table 1 provides a short description of each category. These categories were created two separate experts who viewed the video, created a timeline of the video content (both on screen and spoken by the instructor), then compared these timelines for overlap and discrepancies. Through repeated watching, discussion, and refinement, these timeline categories were iteratively reduced to six broad categories: (1) introduction, (2) python, (3) parameter values, (4) physics (5) physics/python discussion, (6) lab/python discussion. Table 1 contains short descriptions of each category.

**TABLE 1.** Video Timeline categories describe the color codes seen in Figure 2.

| Category | Description |
|---|---|
| Introduction (black) | High level introduction to video material |
| Python (yellow) | Discussion of programming concepts |
| Parameter Values (red) | The video gives very specific information for parameter assignment or specific code necessary to complete the assignment |
| Physics (blue) | Discussion of physics concepts |
| Physics/Python Discussion (cyan) | Discussion of the connection between Python code and the physics it is modeling |
| Lab/Python Discussion (magenta) | Discussion of the connection between Python code and the observed data |

The *introduction* in this video spans the first 1.5 minutes and presents the merits of including computation in an introductory physics course. *Python* sections discuss programming concepts (e.g. describing the difference between assignment and the equals sign). *Parameter values* represent a section of the video that focuses on specific parameters (in this video, mass, initial velocity, initial position, etc.) required to complete the assignment. The *physics* category exclusively discusses physics (e.g., discussing why the net force is zero for a ball with constant velocity). The *physics/python discussion* attempts to explain the connection between python code and the physics it represents. *Lab/Python discussion* constitutes portions of the video that compare simulated data to experimental data.

While there are a variety of categories present in this video, students typically attended to parameter values. That is, in Figure 2, black peaks appear to correspond to red bars. The pause peak that appears between 404 seconds and 418 seconds corresponds to the time when the mass of the ball is written on the screen. It is followed by a short (~20 seconds) comparison between how students would write the mass algebraically compared to how they would write it in Python. Eighty-one students (50.3%) paused here. The seek peaks that appear before this pause peak coincide with the beginning of the discussion on mass.

The second pause peak appeared between 517 seconds and 548 seconds when initial conditions (initial position and initial velocity) for the ball were written on screen. 103 students (64.0% of viewers) paused here. Again we observed a seek peak coinciding with a pause peak. In this portion of the video, students were instructed to run their code over a given time relevant to their observations. The third pause peak between 755 seconds and 776 seconds appeared when this limit for the iteration was given.

We conclude that students used elements of this video as a reference to complete the given assignment. This may indicate that students focused on finding "correct" solutions via information in the video. 90 students (55.9% of viewers) paused here. The observed pausing behavior indicates that students used the video to find the appropriate parameter values to get their code running.

The fourth pause peak between 825 seconds and 835 seconds immediately after the third interaction point requires further explanation. The third interaction point asked the students to "guess" what the correct code for implementing Newton's second law would look like. It presented four different versions of how to implement Newton's second law in Python (only one was correct), and prompted students to select one. Immediately afterward, the answer was given. The fourth pause peak suggests that some students paused the video so that they could copy down the correct version of the Newton's $2^{nd}$ law update.

## CONCLUSIONS

Clickstream data can provide valuable feedback about time on task, where students focus their attention, and what should be focused on during class. Students who watched this video seem to focus on elements that facilitated a "correct" solution. This indicates that students may not be attending to conceptual elements present in the video any more than they would in a passive lecture.

While the results presented in this paper are promising, they represent the analysis of only one out of the seventy-eight videos students watched in this semester. Furthermore, the interaction with this video could be unique to this type (lab) of video. Future work will apply these timeline categories (and more if necessary) to other videos.

This work was supported by the Bill and Melinda Gates Foundation and the Georgia Governor's Office of Student Achievement.